\documentstyle[graphicx]{mn}
\newcommand{\ls}
 {\mathrel{\hbox{\rlap{\hbox{\lower4pt\hbox{$\sim$}}}\hbox{$<$}}}}
\newcommand{\gs}
 {\mathrel{\hbox{\rlap{\hbox{\lower4pt\hbox{$\sim$}}}\hbox{$>$}}}}

\newcommand{\et}{et al.\ }
\newcommand{\rosat}{{\it ROSAT}}

\newcommand{\asca}{{\it ASCA}}

\title[Spectral Complexity in NLS1s]
	{X-ray spectral complexity in narrow-line Seyfert 1 galaxies}
\author[S.\ Vaughan \et]
	{S.\ Vaughan$^{1}$, J.\ Reeves$^{1}$, R.\ Warwick$^{1}$, R.\ Edelson$^{1,2}$\\  
$^1$X-Ray Astronomy Group; Department of Physics and Astronomy; Leicester 
	University; Leicester LE1 7RH; U.K.\\
$^2$Department of Physics and Astronomy; University of California, 
	Los Angeles; Los Angeles, CA 90095-1562; U.S.A.\\
}

\date{Submitted 30 March 1999; Revised version 11 May 1999; Accepted 20
May}

\pagerange{\pageref{firstpage}--\pageref{lastpage}}
\pubyear{1999}

\begin{document}

\maketitle

\label{firstpage}

\begin{abstract}

We present a systematic analysis of the X-ray spectral properties of 
a sample of 22 ``narrow-line'' Seyfert~1 galaxies for which data are available
from the ASCA public archive. Many of these sources, which were selected on the 
basis of their relatively narrow H$\beta$ line width (FWHM~$\leq$2000~km/s),
show significant spectral complexity in the X-ray band.
Their measured hard power-law continua have photon indices spanning 
the range $1.6 - 2.5 $ with a mean of 2.1, which is only slightly steeper 
than the norm for ``broad-line'' Seyfert~1s.  All but four of the sources 
exhibit a soft excess, which can be modelled as blackbody emission 
($T_{bb} \approx 100-300$~eV) superposed on the underlying power-law. 
This soft component is often so strong that, even in the relatively hard 
bandpass of ASCA, it contains a significant fraction, if not the bulk, 
of the X-ray luminosity, apparently ruling out models in which the soft
excess is produced entirely through reprocessing of the hard continuum.

Most notably, 6 of the 22 objects show evidence for a broad absorption 
feature centred in the energy range 1.1--1.4~keV , which could be
the signature of resonance absorption in highly ionized material. A further
3 sources exhibit ``warm absorption'' edges in the 0.7--0.9~keV bandpass. 
Remarkably, all 9 ``absorbed'' sources have H$\beta$ line widths below 
1000~km/s, which is less than the median value for the 
sample taken as a whole. This tendency for very narrow line widths to 
correlate with the presence of ionized absorption features in the soft X-ray 
spectra of NLS1s, if confirmed in larger samples, may provide a further clue 
in the puzzle of active galactic nuclei.

\end{abstract}


\section{Introduction}

Significant hard X-ray luminosity is now regarded as one of the defining 
characteristics of Seyfert 1 galaxies. X-ray spectral and variability 
observations establish that the emission is produced in the innermost regions 
of the Seyfert nucleus and provide a means of investigating both the nuclear 
geometry and the processes accompanying accretion onto a supermassive 
black-hole (e.g. Mushotzky, Done \& Pounds 1993).

The hard X-ray spectra of normal Seyfert~1s can, to first order, 
be described in terms of a power-law continuum with a photon index of typically
$\Gamma = 1.9$ (Nandra \& Pounds 1994). In many sources additional
spectral features are observed, such as a spectral turn-up above 10~keV 
and strong iron K$\alpha$ fluorescence in the 6 - 7 keV band, which 
can be attributed to Compton reflection  of the continuum by optically thick 
matter, possibly in the form of an accretion disk (e.g. George \& Fabian 1991; 
Matt, Perola \& Piro 1991). The presence of a soft X-ray excess may in 
turn represent the high energy tail of the thermal emission generated in the 
inner regions of the disc (e.g. Czerny \& Elvis 1987). Recent interest has 
also focussed on the detection of ``warm absorption'' features 
in the soft X-ray spectra  of Seyfert 1 galaxies which are an imprint of 
highly photoionized material lying in the line of sight to the nuclear source 
(Reynolds 1997; George et al. 1998).

Narrow-line Seyfert 1s (NLS1s) represent a subclass of objects occupying one 
extreme of the measured range of optical line widths, {\it i.e.} H$\beta$
FWHM $\leq 2000$ km/s (Osterbrock \& Pogge 1985). 
Observations with \rosat\ have shown that NLS1s frequently have stronger 
soft excesses and increased variability in the 0.1--2~keV band than 
more normal, ``broad-line'' Seyfert~1s (hereafter BLS1s; 
Boller, Brandt \& Fink 1996). 
A possible explanation of the extreme properties of NLS1s is that these 
are active galactic nuclei (AGN)  containing black holes of relatively modest
mass which, nevertheless, are  accreting at a high rate 
(Pounds, Done \& Osborne 1995). 
In this model, the lower mass black hole gives rise to smaller Keplerian 
velocities of the broad-line region clouds while 
the high accretion rate yields comparable luminosity to that of larger
mass black holes and leads to increased disc emission and an enhanced 
soft excess (Ross, Fabian \& Mineshige 1992). In some NLS1s there is
tentative evidence
for reflection from a highly ionized disc (e.g.
Comastri \et\ 1998), which fits naturally with the high accretion rate model, 
as the disc surface is expected to be highly ionized (Matt, Fabian \& Ross 
1993).  Recent studies suggest that
NLS1s, in addition to their extreme soft X-ray spectra, also exhibit
steeper intrinsic hard X-ray continua than their broad-line counterparts
(e.g. Brandt, Mathur \& Elvis 1997), an effect which could arise, for
example, if the strong soft excess effectively cools the accretion disk 
corona in which the underlying power-law is formed (Pounds, Done \& Osborne
1995; Maraschi \& Haardt 1997)

Our objective in this paper is to investigate the X-ray spectral 
properties of a sample of NLS1s. Specifically we are interested in
whether there are any further X-ray spectral characteristics of NLS1s,
over and above those noted earlier, which might be described as distinctive 
of this class of object.  In practical terms we have carried out a 
systematic analysis of 24 observations of 22 NLS1s taken from the \asca\
public data archive.  The remainder of this paper is organised as follows.
In the next section we define the source sample and detail the observations 
and data reduction techniques. In Section~3 we go on to describe the 
results of our X-ray spectral analysis and in Section~4 present a
discussion of these results. Finally we summarise our
conclusions in Section~5.

\begin{table*}
\centering
\caption{The \asca\ NLS1 sample. The columns contain the following
information: (1) The source name; (2) The sequence
number of the observation; (3) \& (4) The observation pointing
position;  (5) The exposure time for SIS--0 (ks); (6) The line-of-sight 
Galactic hydrogen column density from Dickey \& Lockman (1990) (in units of 
$10^{20}$ cm$^{-2}$); (7) The source redshift; (8) The H$\beta$ line 
width (km/s); (9) References to the information in columns (7) \& (8).}

\begin{tabular}{@{}lcccccccc@{}}
Name  & \asca\ & RA       & Dec    & SIS-0 & $N_{H}$ & z &H$\beta$
FWHM & Ref.\\
 & Sequence & (J2000)& (J2000)& Exp. &  & & & \\  
  (1) & (2) & (3) & (4) & (5) & (6) & (7) & (8) & (9)\\
 & & & & & & & & \\
Mkn 335       & 71010000 &  00 06 04 &  20 11 12 & 18 & 3.99 & 0.026 & 1640&  1 \\
I Zw 1        & 73042000 &  00 53 41 &  12 46 40 & 27 & 4.99 & 0.061 & 1240&  2 \\
Ton S180      & 74081000 &  00 57 28 & --22 17 50& 44 & 1.55 & 0.062 & 1000&  2 \\
PHL 1092      & 75042000 &  01 40 01 &  06 24 32 & 68 & 4.07 & 0.396 & 1300&  2 \\
RX J0148--27   & 75048000 &  01 48 30 & --27 53 19& 34 & 1.42 & 0.121 & 1050&  3 \\
NAB 0205+024  & 74071000 &  02 07 44 &  02 38 02 & 41 & 3.51 & 0.155 & 1050&  4 \\
RX J0439--45   & 75050000 &  04 40 14 & --45 39 46& 43 & 2.02 & 0.224 & 1010&  3 \\
PKS 0558--504  & 74096000 &  06 00 04 & --50 22 06& 33 & 4.39 & 0.137 & 1500&  2 \\
1H 0707--495   & 73043000 &  07 08 26 & --49 37 47& 34 & 5.80 & 0.040 & 1000&  2 \\
RE J1034+39 (1)& 72020000 & 10 35 04 &  39 40 28 & 28 & 1.02 & 0.042 & 1500&  2 \\
RE J1034+39 (2)& 72020010 & 10 34 14 &  39 36 05 & 11 & 1.02 & 0.042 & 1500&  2 \\
Mkn 42        & 75056000 &  11 53 12 &  46 11 26 & 36 & 1.99 & 0.024 & 670 &  2 \\
NGC 4051 (1)  & 70001000 &  12 03 06 &  44 32 33 & 25 & 1.32 & 0.002 & 990 &  2 \\
NGC 4051 (2)  & 72001000 &  12 02 43 &  44 29 50 & 69 & 1.32 & 0.002 & 990 &  2 \\
PG 1211+143   & 70025000 &  12 14 13 &  14 03 32 & 13 & 2.74 & 0.080 & 1860&  1 \\
PG 1244+026   & 74070000 &  12 46 17 &  02 19 24 & 37 & 1.75 & 0.048 & 830 &  1 \\
IRAS 13224--3809& 72011000 & 13 24 52 & --38 25 57& 82 & 4.79 & 0.067 & 650 &  2 \\ 
PG 1404+226   & 72021000 &  14 06 06 &  22 22 50 & 34 & 2.14 & 0.098 & 880 &  1 \\
Mkn 478       & 73067000 &  14 41 43 &  35 24 57 & 29 & 1.03 & 0.079 & 1450&  2 \\
PG 1543+489   & 75059000 &  15 46 00 &  48 48 06 & 39 & 1.59 & 0.400 & 1560&  1 \\
IRAS 17020+454& 73047000 &  17 03 09 &  45 36 52 & 35 & 2.22 & 0.060 & 490 &  2 \\
Mkn 507       & 74033000 &  17 49 00 &  68 37 26 & 32 & 4.38 & 0.056 & 960 &  2 \\
IRAS 20181--224& 73075000 &  20 20 53 & --22 40 10& 51 & 5.96 & 0.185 & 580 &  2 \\
Ark 564       & 74052000 &  22 42 34 &  29 38 21 & 47 & 6.40 & 0.024 & 750 &  2 \\

\end{tabular}

\medskip

\raggedright

REFERENCES: (1) Wang \et (1996). (2) Brandt (1996). 
(3) Grupe (1996). (4) Zheng \& O'Brien (1990).

\end{table*}

\section{ Observations and data reduction }

We have selected 22 objects, identified in the literature as Seyfert~1 
galaxies or quasars with H$\beta$ FWHM $\leq$2000 km/s, for which \asca\ 
data were available at the time of the analysis (late 1998). 
The sample, given in Table~1, is not complete in any sense except that it 
consists of all suitable spectra of NLS1s in the \asca\ public archive.
The redshift distribution of the sample is shown in Fig.~1.
The NLS1 galaxy Mkn~957 was also observed by \asca\ but these
data have not been analysed due to the very short exposure time of the
observation ($\sim$3~ks). In addition data on Kaz~163 have been excluded; 
Kaz~163 is in the same GIS field as Mkn~507 but its image falls
at the extreme edge of the SIS detector, making the data unsuitable
for detailed spectral analysis.
 
\begin{figure}
 \begin{center}
\rotatebox{-90}{\includegraphics[width=6.5cm]{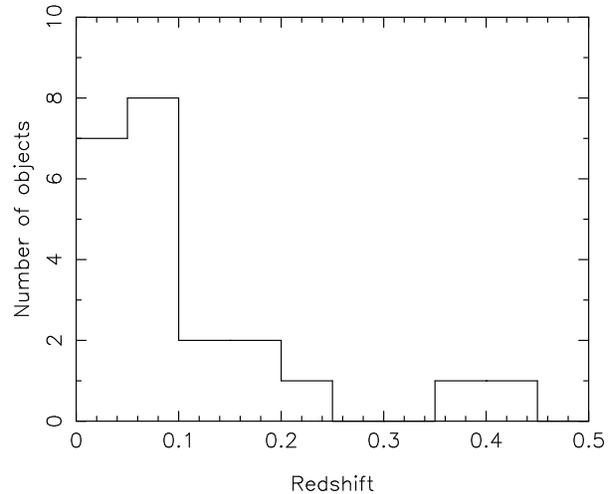}}
 \end{center}
 \caption{The redshift distribution of the NLS1 sample. }
\end{figure}

\asca\ carries four instruments which are operated simultaneously
namely two solid-state imaging spectrometers 
(SIS-0 and SIS-1; Burke \et 1991) and two gas imaging spectrometers
(GIS-2 and GIS-3; Kohmura \et 1993). In the current analysis,
standard data screening criteria were  applied to all the data.
Table~1 lists the resulting exposure times (specifically for SIS-0,
although data from all four \asca\ instruments are utilised here). 
Since we are concerned only with the spectral properties  of the sources,
time-averaged spectra have been extracted from each observation. Counts were 
accumulated, where possible, in circular apertures of 3--4 
arcmin radius centred on the source. The background was estimated using 
source free regions from the same observation and, in the case
of the GIS observations, at a similar off-axis angle to the source. 
After background subtraction the pulse height spectra were binned to give 
at least 20 counts per spectral channel. In the subsequent analysis 
the SIS data were restricted to the 0.6--10~keV energy range and
GIS data to the 0.8--10~keV band so as to reduce the possible impact of 
calibration uncertainties.

As a check of the possible impact of spectral variability on the time-averaged
source spectra, we extracted the 2--10 keV and 0.5--2.0 keV light curves
of each source (SIS-0 and SIS-1 co-added) and binned these on the satellite 
orbital timescale (5760 s). We then used a $\chi^{2}$ test to check the 
constancy of the spectral hardness (defined as ratio of the 2--10 keV count 
rate to the 0.5--2.0 keV count rate) during the observation. Five out of 22 
sources gave unacceptable values of $\chi^{2}$ at the 95\% level 
(NAB 10205+024, RX J0439--45, NGC 4051, Mkn 478, Ark 564) indicative of 
underlying spectral variability. However, with the exception of NGC 4051 
there was little evidence that these variations corresponded to any 
clear spectral trends or to a systematic effect (e.g. a spectral hardness 
versus flux correlation). We conclude that the implicit assumption of the 
analysis 
which follows, namely that the observation-averaged spectra provide a good
representation of the source properties, is a reasonable one.

\section{X-ray spectral analysis}

Spectral fitting of the binned source spectra was carried out using the 
\textsc{xspec v10.0} software package. In general, data from all four
detectors have been fitted simultaneously, but with the relative
normalisations of the spectra free to vary between the four instruments.
The quoted errors on the derived best-fitting model parameters correspond
to a 90\% confidence level for one interesting parameter
({\it i.e.} a $\Delta \chi^{2}=2.7$ criterion) unless otherwise stated.
Values of $ H_0 = 50 $~km/s/Mpc and $ q_0 = 0.5 $ are assumed
throughout this section and fit parameters (specifically
the emission and absorption line energies) are quoted for 
the rest frame of the source.

\subsection{Initial spectral fits}

Broadly, our strategy has been to apply relatively simple spectral models, 
so as to characterize the properties of the whole sample of sources. 
Thus initially a model comprising just a power-law continuum with 
absorption due to the line-of-sight Galactic $N_H$ (see Table 1) 
was employed as a way of illustrating the variety of X-ray spectra
exhibited by NLS1 galaxies. Fig.~2 shows the residuals from these fits 
for all 24 observations, which in most instances have unacceptable 
minimum $\chi^{2}$ values. Cursory inspection of Fig.~2 shows that soft 
excesses are often present below $\sim$2~keV and that broad absorption
and/or emission complexes are also evident in a number of the sources.
In general, an iron K$\alpha$ emission-line is difficult to pick out 
(by eye), although NGC 4051 provides the exception.
In one source, Mkn~507 the largest residual occurs in the lowest spectral 
channel suggesting the possibility of cold absorption over and above the 
Galactic column density. In this case we obtain an improved fit
($\Delta \chi^{2}=12$) when the $N_H$ is allowed to be a free parameter,
the specific value derived being 
$N_{H} = 3.2 \pm 1.5 \times 10^{21} \rm~cm^{-2}$
(see Iwasawa, Brandt \& Fabian 1998). 
In the rest of this paper we assume that Mkn~507 is the only NLS1 in the 
sample with excess cold absorption.

\subsection{The 2--10~keV spectral form}

In order to characterize better the hard X-ray spectrum of each NLS1, we 
next excluded all the spectral data below 2~keV, where evidently soft 
excesses and warm absorbers add to the spectral complexity. The 2--10~keV 
spectra were again fitted with a simple power law plus  Galactic absorption 
model but with the addition of a narrow  Gaussian 
emission line ($\sigma=0.01$~keV) to represent iron K$\alpha$ emission 
in the 6.4--7.0 keV bandpass. Table~2 summarizes the results. 
Acceptable fits were found for most sources, with only three instances 
of (reduced) $\chi^{2}_{\nu} \geq 1.1$. We note that the derived photon 
indices have a mean value of 2.1 with a standard deviation of 0.3.

For nine objects the inclusion of the iron-line component is merited
in terms of the resulting improvement in $\chi^{2}$ ({\it i.e.}
$\Delta\chi^{2} > 6.2$  implying better than
$95$\% confidence in the F-test for two additional parameters).
Also two further sources show at least a marginal improvement 
in $\chi^{2}$ (see Table 2). There is some evidence for lines 
originating from fairly highly ionized iron species 
({\it i.e.} $>$Fe\textsc{xx} and above) in I~Zw~1, Ton~S180, PKS~0558-504, 
PG~1244+026 and Ark~564. The measured line equivalent widths
range from 100--600 eV, albeit with large uncertainties.
Similarly the constraints on the equivalent width of a (neutral) 
iron line in those sources lacking a significant line detection are 
generally rather weak.  In general the signal/noise ratio was too
poor to meaningfully constrain the intrinsic line widths. However,
the second observation of NGC~4051 provides the exception since in this
case there is evidence for a broadened iron K$\alpha$ feature. The best fit
parameters for the line are E$=6.29\pm0.09$~keV, intrinsic width
$\sigma=0.36^{+0.23}_{-0.15}$~keV and EW$=240^{+90}_{-80}$~eV. The
improvement in the fit upon adding this broad feature is
$\Delta\chi^{2}=76.8$. An attempt has been made to fit the line with
the relativistic ``diskline'' profile of Fabian \et (1989);
we find a mildly ionized line (E$=6.6^{+0.1}_{-0.2}$~keV),
with a near to face-on inclination ($ i < 17^{\circ}$) gives the 
best fit to the data (assuming values for R$_{in}$, R$_{out}$ 
and emissivity index of 10GM/c$^{2}$, 100GM/c$^{2}$ and --2, respectively).
Note that Guainazzi \et (1996) discuss the the spectral fitting of this \asca\
observation of NGC~4051 in some detail.

\begin{table*}
\centering
\caption{ The results of the 2--10~keV spectral fitting. The columns give 
the following information: (1) The source name; (2) The power-law photon index;
(3) The 2-10 keV flux ($10^{-13}$~erg cm$^{-2}$ s$^{-1}$); 
(4) The 2-10 keV luminosity ($10^{43} \rm~erg~s^{-1}$);
(5) The iron line energy (keV); (6) The iron-line equivalent width (eV);
(7) The improvement in the fit when the iron line is
included; (8) The {\it reduced} $\chi^{2}$ and number of degrees of freedom 
for the best fitting model. }

\begin{tabular}{@{}lccccccc@{}}
               & $\Gamma_{2-10}$ & Flux$_{2-10}$ & L$_{2-10}$ &
E$_{Fe}$ & EW$_{Fe}$ & $\Delta \chi^{2}_{Fe}$ & $\chi^{2}_{\nu}$/dof \\
  (1)          & (2) & (3) & (4) & (5) & (6) & (7) & (8) \\
               &  & & & & & & \\
Mkn 335        & $1.89\pm0.05$ & 109.0& 3.2 & $6.41\pm0.08$ &
$193\pm78$ & 16.3 & 1.00/458\\     
I Zw 1         & $2.27\pm0.10$ & 34  & 5.9  & $6.73^{+0.15}_{-0.30}$ & $450^{+250}_{-225}$ & 11.0 &1.04/262\\ 
Ton S180       & $2.39\pm0.05$ & 45.3& 8.3  & $6.65\pm0.10$ & $180\pm90$ & 10.9 &1.05/591 \\
PHL 1092       & $1.67\pm0.30$ & 3.6 & 32.0 & -- & $<490$ & -- &1.18/132\\
RX J0148--27   & $1.99\pm0.17$ & 8.5 & 5.7  & $6.5\pm0.2$ & $670\pm300$ & 8 &0.86/87\\ 
NAB 0205+024   & $2.10\pm0.09$ & 26.8& 30   & -- &$<122$  & --&0.85/327\\
RX J0439--45   & $2.25\pm0.16$ & 10.3& 22.0 & -- & $<220$ & -- &1.09/149\\
PKS 0558--504  & $2.25\pm0.04$ & 121 & 110  & $6.65^{+0.35}_{-0.20}$ & $65\pm40$ & 6.5 &0.98/820\\
1H 0707--495   & $2.40\pm0.20$ & 9.6 & 0.7  & -- & $<375$ & -- &1.04/114\\
RE J1034+39 (1)& $2.46\pm0.23$ & 10.2& 0.7  & -- & $<490$ & -- &1.08/98\\
RE J1034+39 (2)& $2.73\pm0.40$ & 8.6 & 0.7  & -- &$<1560$  & -- &0.97/32\\   
Mkn 42         & $2.01\pm0.17$ & 10.5& 0.3  & -- & $<380$ & -- &0.93/122\\   
NGC 4051 (1)   & $1.91\pm0.03$ & 291 & 0.045& $6.43\pm0.04$ & $145\pm40$ & 35.6 &1.00/905\\  
NGC 4051 (2)   & $1.84\pm0.02$ & 234 & 0.040& $6.37\pm0.03$ & $127\pm25$ & 66.5 &1.06/1449\\   
PG 1211+143    & $2.03\pm0.10$ & 34  & 9.7  & $6.38^{+0.12}_{-0.20}$ & $260\pm130$ & 10.7 &0.94/342\\
PG 1244+026    & $2.35\pm0.10$ & 26.8& 2.8  & $7.0\pm0.1$ & $460\pm210$ & 11 &0.99/255\\
IRAS 13224--3809& $1.55\pm0.22$& 5.16& 1.0  & -- & $<290$ & -- &1.12/182\\
PG 1404+226    & $1.6\pm0.4$   & 6.4 & 2.4  & --& $<1220$ & -- &0.84/57\\
Mkn 478        & $1.92\pm0.10$ & 25  & 7.0  & $6.38^{+0.18}_{-0.10}$ & $200\pm135$ & 5.7 &1.02/216\\
PG 1543+489    & $2.46\pm0.32$ & 4.0 & 37.2 & -- & $<350$ & -- &0.87/76\\
IRAS 17020+454 & $2.20\pm0.06$ & 78.4& 12.6 & $6.5\pm0.5$ & $95\pm70$
& 5 & 0.87/577\\
Mkn 507        & $1.61\pm0.30$ & 6.7 & 9.3  & -- & $<715$ & -- &1.02/50\\
IRAS 20181--224& $2.33\pm0.12$ & 10.4& 17.5 & -- & $<120$ & -- &1.14/185\\
Ark 564        & $2.50\pm0.03$ & 205.7& 5.2 & $7.0\pm0.3$ & $123\pm50$
& 15 &0.99/1171\\

\end{tabular}

\end{table*}

\subsection{The soft X-ray spectra}

Extrapolation of the best-fit 2--10~keV spectrum (as defined in
Table~2) down to 0.6~keV in most cases results in a poor fit of 
the soft X-ray spectrum, with the most common residual feature being
an excess of soft flux. In the spectral fitting we have attempted to
match this soft excess with an additional continuum component.
Specifically we use a single blackbody component, although a second power law 
often provides an equally good fit. (In the latter case
the second power-law is typically steeper than the first by 
$\Delta \Gamma \simeq 0.5$ with a break energy in the range 1--2~keV).
Table~3 summarises the results of fitting a power law plus 
blackbody model (note that from here on the iron line parameters are 
frozen at the values obtained in the earlier 2--10 keV fits).
In all but four objects (I~Zw~1, PG~1543+489, Mkn~507 and
IRAS~20181--224) a soft excess component provides a significant
improvement in the fit, demonstrating that soft excesses are 
a very common feature in NLS1s. The underlying power-law photon
indices given in  Table~3 have a mean of 2.12 and a standard deviation
of 0.26. The fact the mean is very similar to that
obtained earlier for 2--10~keV fits demonstrates that 
the latter fits are relatively immune to the presence of the soft excess.
 
\begin{table*}
\centering
\caption{The full 0.6--10 keV spectral fits. The columns give the following
information: (1) The source name; (2) The power-law photon index; (3) 
The blackbody temperature (eV); (4) The 0.6--10~keV flux 
($10^{-13}\ \rm{erg}\ \rm{cm}^{-2}\ \rm{s}^{-1}$); (5) The derived 
unabsorbed luminosity (10$^{43} \rm~erg~s^{-1}$); (6) The ratio of 
the  0.6--10~keV luminosities of the blackbody and power-law  
components; (7) The improvement in the fit when the blackbody component 
is added; (8) The {\it reduced} $\chi^{2}$ and number of degrees of
freedom for the best fit model.}

\begin{tabular}{@{}lccccccc@{}}
Name          & $\Gamma$ & kT & Flux$_{0.6-10}$ &
L$_{0.6-10}$  & L$_{bb}$/L$_{pl}$ & $\Delta \chi^{2}$ & $\chi^{2}_{\nu}$/dof\\  
  (1)         & (2) & (3) & (4) & (5) & (6) & (7) & (8) \\
              & & & & & & & \\
Mkn 335       & $1.96\pm0.04$ & $160\pm14$ & 191  & 5.7   & 0.12 & 125 & 1.01/757\\
I Zw 1        & $2.30\pm0.03$ & 150$^{f}$       & 72  & 14.0   & $<0.01$ & --& 0.94/594\\
Ton S180      & $2.36\pm0.05$ & $190\pm5$  & 137 & 24.7   & 0.29 & 226 &1.10/988\\
PHL 1092      & $1.67\pm0.11$ & $162\pm15$ & 7.0   & 73   & 0.80 & 112 &1.12/268\\
RX J0148--27  & $2.15\pm0.15$ & $129^{+32}_{-124}$ & 18.3   & 12.7 & 0.15 & 12 &0.91/230\\
NAB 0205+024  & $2.10\pm0.08$ & $218\pm19$ & 53  & 63   & 0.16 & 36 & 0.98/697\\
RX J0439--45  & $2.28\pm0.10$ & $137\pm11$ & 27.4  & 81   & 0.58 & 152&1.02/379\\
PKS 0558--504 & $2.26\pm0.03$ & $230\pm15$ & 257 & 247  & 0.08 & 34 &1.03/1217\\
1H 0707--495  & $2.33\pm0.10$ & $107\pm4$  & 39.6  & 3.7   & 1.21 & 429 &1.29/338\\
RE J1034+39 (1)&$2.35\pm0.13$ & $131\pm9$  & 32.8  & 2.7   & 0.64& 104  &0.95/298\\
RE J1034+39 (2)&$2.45\pm0.20$ & $142\pm20$ & 33.6  & 2.8   & 0.49 & 23  &0.78/125\\
Mkn 42        & $1.89^{+0.17}_{-0.09}$     & $227\pm40$     & 20.6 & 0.53 & 0.15 & 8 &0.99/271\\
NGC 4051 (1)  & $1.97\pm0.02$ & $86\pm5$   & 561 & 0.10 & 0.22 & 824 &1.04/1205\\
NGC 4051 (2)  & $1.94\pm0.01$ & $106\pm5$  & 402 & 0.07 & 0.11 & 1160 &1.13/1846\\
PG 1211+143   & $2.07\pm0.06$ & $110\pm12$ & 72  & 23.0   & 0.17 & 57 &0.95/713\\
PG 1244+026   & $2.31\pm0.08$ & $210\pm10$ & 75  & 8.1    & 0.26 & 64 &1.17/535\\
IRAS 13224--3809&$1.60\pm0.13$& $119\pm4$  & 18  & 4.7   & 2.1 & 537&1.28/426\\ 
PG 1404+226   & $1.87\pm0.20$ & $115\pm7$  & 12.9  & 9.1   & 1.9 & 177 &1.20/167\\
Mkn 478       & $1.96\pm0.05$ & $89\pm12$  & 50  & 14.5   & 0.23 & 101 &1.03/495\\
PG 1543+489   & $2.48\pm0.20$ & $436\pm436$& 10.7  & 105  & 0.04 & 1&0.91/206\\
IRAS 17020+454& $2.21\pm0.05$ & $260\pm36$ & 148 & 29   & 0.05 & 9 &1.06/972\\
Mkn 507       & $1.74\pm0.24$ & 150$^{f}$       & 8.6   & 1.5    & $<0.08$& -- &0.96/97\\
IRAS 20181--224&$2.33\pm0.11$ & $258^{+62}_{-79}$ & 19.3 & 47.4  & 0.08 & 3 & 1.15/389\\ 
Ark 564       & $2.44\pm0.03$ & $195\pm4$ & 568 & 16.7 & 0.21 & 558 & 1.27/1567\\
\end{tabular}
\end{table*}

Nine objects show signs of additional spectral complexity below 2~keV
even after the inclusion of the blackbody component in the fit (Table~3).
Unfortunately modelling of the \asca\ spectra in terms of additional soft 
X-ray features is not particularly straight-forward. Radiation damage to the 
CCDs and other factors have meant that there is increasing uncertainty in 
the calibration of the detectors below 1~keV and, in particular, 
it has been noted that the two SIS
instruments often give divergent spectra at the lowest
energies (although these calibration uncertainties only dominate over
uncertainties in the background subtraction for relatively bright sources).
Also since most NLS1 galaxies appear to exhibit a soft X-ray excess 
it is difficult to distinguish subtleties in the form of the soft continuum
from the effects of putative broad emission and/or absorption features.

Recent studies of the X-ray spectra of NLS1s  
(e.g. Leighly \et 1997b; Fiore \et 1998) have established
that in addition to classical ``warm-absorption'' features, NLS1s
often show anomalous absorption features in the 1--2 keV band.
Both types of absorption are spectrally complex
and will merit more detailed analysis using the predictions
of appropriate photoionization codes, once high sensitivity 
X-ray data with good spectral resolution become available from missions such as 
AXAF, XMM and ASTRO-E (e.g. Nicastro et al. 1999). However, for
our present purpose we have taken a very simplistic approach and
have attempted to improve the $\chi^{2}$ in the spectral fits for nine 
sources noted above by including just a {\it single} absorption feature 
(in the form of a broad Gaussian absorption line) in the spectral model.

Table~4 lists for each of the nine sources the line energy, the equivalent 
width and the intrinsic line width obtained when such an absorption feature 
is added to the power-law plus blackbody continuum model. The line energies 
are not consistent with a single value but a bifurcation is suggested, namely
absorption either in the 0.7--0.9~keV range or in the
1.1--1.4~keV range. Absorption features in the former range are usually
interpreted as due to O\textsc{vii} and O\textsc{viii} edges 
indicative of the presence of ionized material along the line of sight. 
The possible origin of the ``anomalous'' absorption features 
observed at $\sim1.2$~keV have recently been discussed by 
Leighly \et (1997b) and Fiore \et (1998).

As noted above the fitting of a Gaussian absorption feature is necessarily
an over simplification of the true picture. In order to investigate the 
individual sources in somewhat more detail we have refitted the spectra of 
the nine objects listed in Table~4 including a variety of additional 
absorption and emission components. For example, we tested for either one or
two absorption edges and also for either a single 
Gaussian emission line or a \textsc{mekal}-type
optically thin thermal spectrum (Kaastra \& Mewe 1993). The outcome was
that in five of the objects an absorption component gave a significantly 
better fit than an emission component. (Of course the modelling of the
underlying soft excess changes  substantially between these two cases 
so as to maintain the match to the observed spectrum in the 0.6--2 keV 
bandpass). Unfortunately the situation is more ambiguous in the remaining 
cases since absorption and emission models can be constructed
which give fairly comparable fits to the \asca\ data. Further comments on
the spectral fitting of the individual nine sources are given in the 
Appendix.

\begin{table}
  \centering
  \caption{X-ray spectral fitting of a Gaussian absorption line.
The columns give the following information: (1) The source name;
(2) The best fit line energy (keV); (3) The line equivalent width (eV);
(4) The intrinsic line width (keV); (5) The improvement in the fit when 
the absorption feature is included (compared to Table 3).}
  \begin{tabular}{@{}lcccc@{}}
Name             & Energy   & EW  & $\sigma$ & $\Delta \chi^{2}$ \\
             (1) & (2) & (3) &    (4)   & (5)  \\
                 &     &     &          &      \\
Ton S180         & $1.22^{+0.07}_{-0.04}$ & $-95\pm16$ & $0.31\pm0.04$ & 20 \\
1H 0707-495      & $1.08\pm0.03$ & $-112^{+25}_{-38}$ & $0.14\pm0.07$ & 32 \\
NGC~4051         & $0.83\pm0.03$ & $-33^{+4}_{-21}$ & $0.15^{+0.15}_{-0.02}$ & 43    \\
PG 1244+026      & $1.31^{+0.07}_{-0.03}$ & $-108\pm50$ & $0.24\pm0.08$ & 24 \\
IRAS 13224-38    & $1.10\pm0.07$ & $-275^{+120}_{-160}$  & $0.24^{+0.06}_{-0.03}$  & 87 \\
PG 1404+226      & $1.16\pm0.05$ & $-100\pm30$ & $0.09\pm0.02$ & 31 \\
IRAS~17020+45    & $0.70^{+0.05}_{-0.02}$ & $-129^{+25}_{-35}$  & $0.12\pm0.05$ & 64 \\
IRAS 20181-22    & $0.90\pm0.03$ & $-42^{+61}_{-24}$ & $<1.6$ & 8 \\
Ark 564          & $1.38\pm0.03$ & $-55^{+24}_{-10}$ & $0.19^{+0.11}_{-0.03}$  & 132 \\
  \end{tabular}
\end{table}

\begin{figure*}
 \begin{center}
\includegraphics[width=14.5 cm]{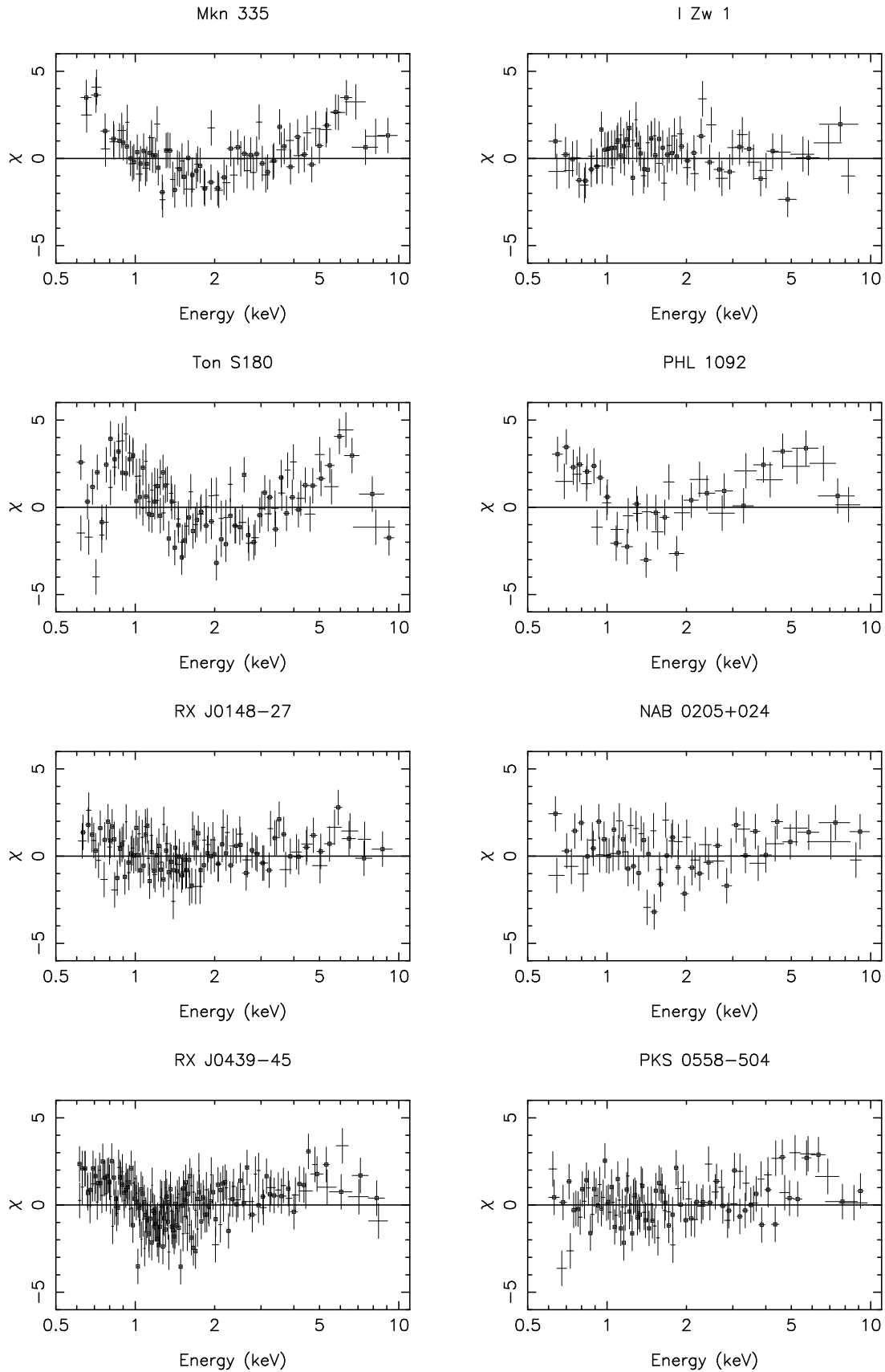}
 \end{center}
 \caption{The residuals from spectral fits to the 0.6--10.0~keV 
\asca\ spectra using a power-law plus Galactic absorption model.
The residuals are in terms of sigmas, with error bars of size one.
For clarity only the SIS data are shown, and the SIS-0 data are marked by
squares. }
\end{figure*}

\begin{figure*}
 \begin{center}
\includegraphics[width=14.5 cm]{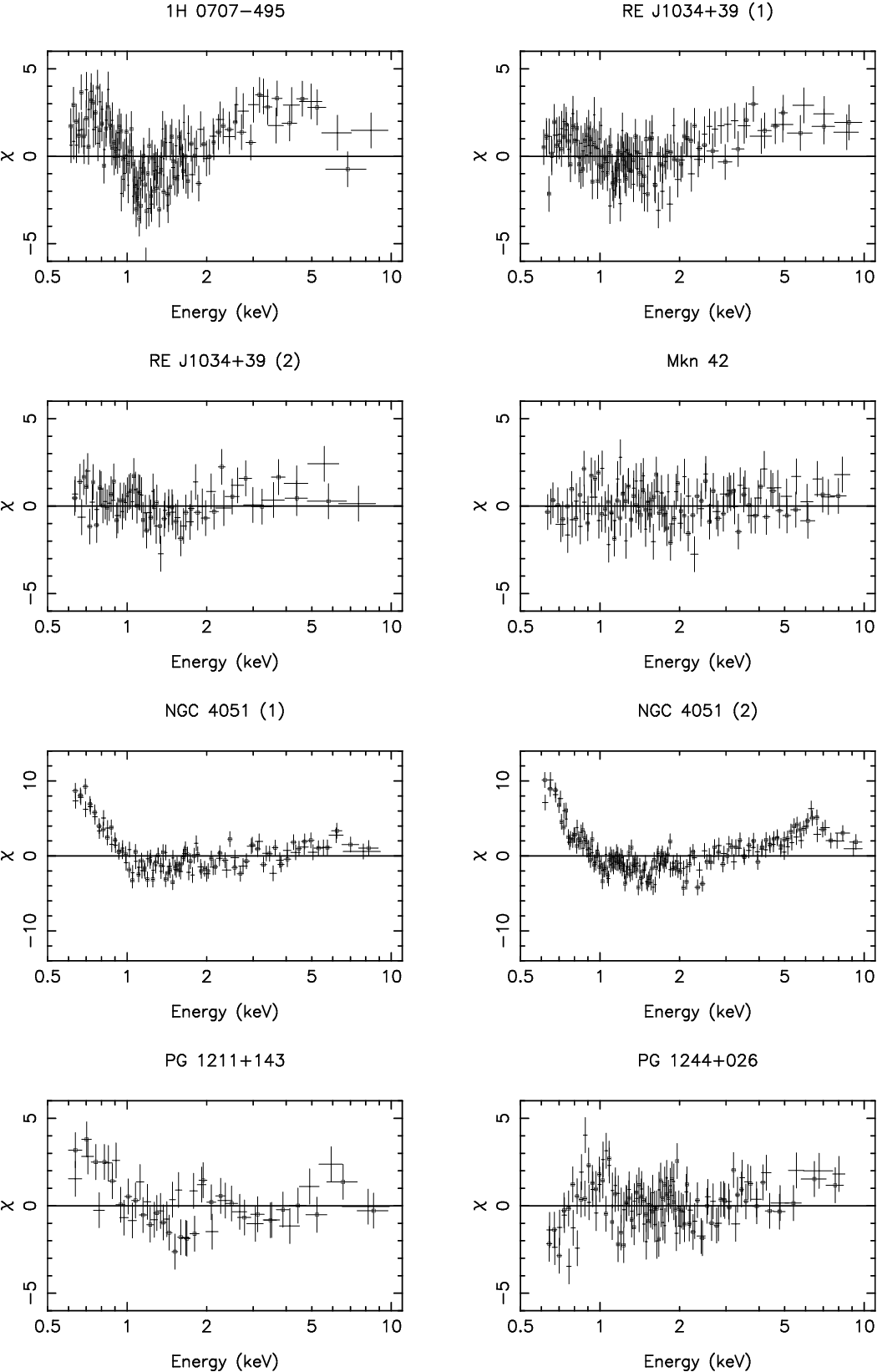}
 \end{center}
\end{figure*}

\begin{figure*}
 \begin{center}
\includegraphics[width=14.5 cm]{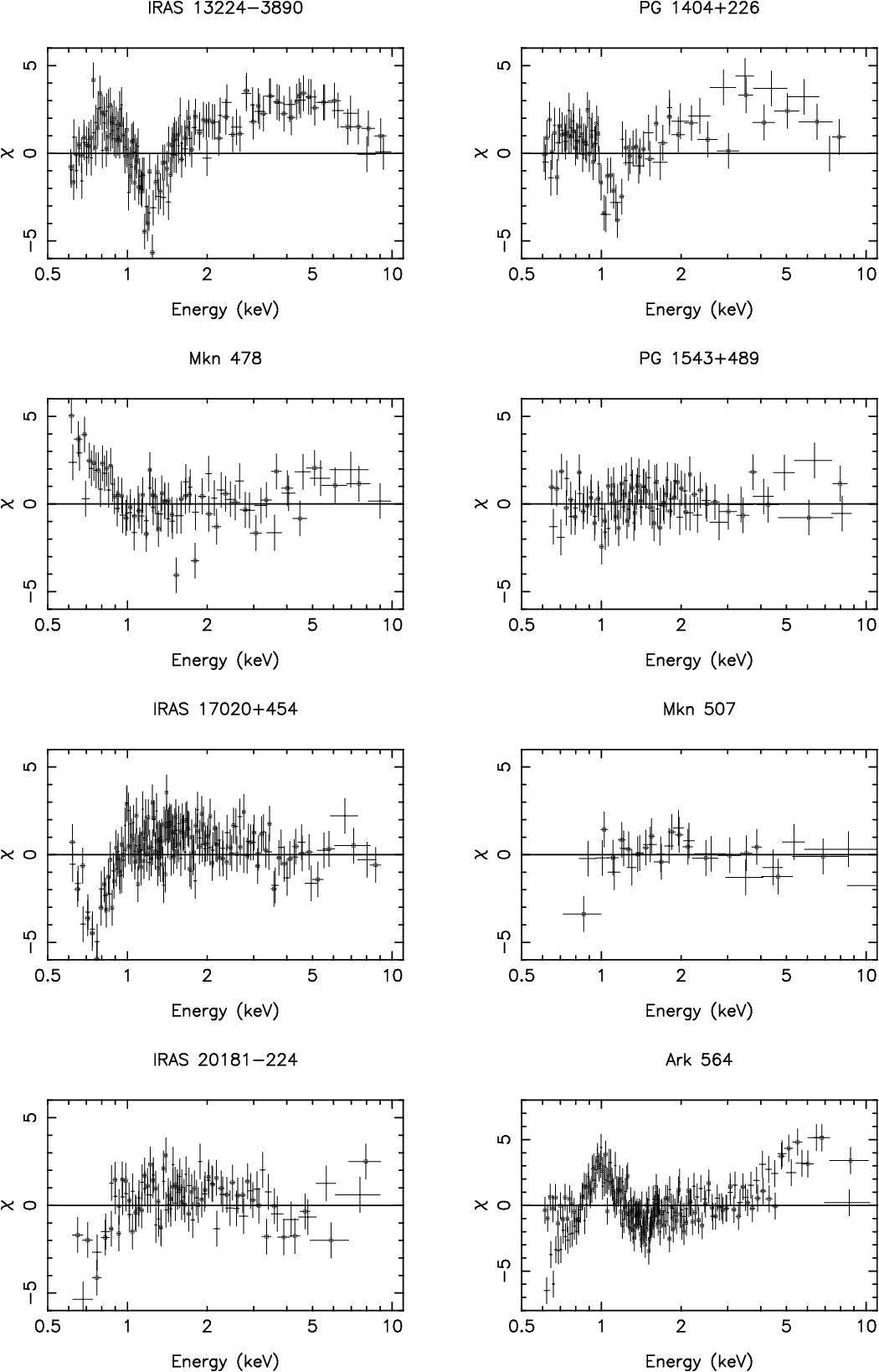}
 \end{center}
\end{figure*}

\setcounter{figure}{2}

\section{Discussion}

\subsection{The X-ray continua of NLS1s}

Using spectral measurements compiled from the published literature
Brandt \et (1997) identify a correlation between the hard power-law
spectral index and the width of the optical H$\beta$ line 
in a sample of Seyfert 1 galaxies, in the sense that NLS1s tend to have 
somewhat steeper hard X-ray spectra than normal broad-line objects.
Fig.~3 shows a plot of power-law photon index (from Table 3)
against the H$\beta$ FWHM line width for our source sample.
Clearly within the restricted range of line width which defines the
NLS1 subclass ({\it i.e.}  H$\beta$ FWHM $\leq 2000 \rm~km~s^{-1}$ )
there is no hint of a correlation.
As noted earlier the mean photon index of our sample of NLS1s
is $\Gamma=2.12$, with a standard deviation of $\sigma$=0.26.
Nandra \et (1997) tabulate the 3--10~keV spectral slopes for 15 BLS1s 
based on \asca\ measurements. These give an average
slope of $\Gamma=1.9$ with a standard deviation of $\sigma=0.17$.
Thus a modest trend in $\Gamma$ over a broad range of H$\beta$ line width
is suggested, consistent with the Brandt \et (1997) result. 
However, we caution that the spectral steepening is a fairly subtle effect 
and the above comparison may not be entirely free of selection bias. For 
example, one could argue that the comparison is of a hard X-ray selected 
sample with one largely selected at soft X-rays (but see the discussion in 
Brandt \et 1997). In counterbalance is the fact that Nandra \et (1997) 
include a reflection component in their analysis, which can have the effect 
of increasing the derived power-law spectral indices by up to 
$\Delta \Gamma \simeq 0.1$. Thus it is possible that magnitude of 
the spectral steepening trend noted above is understated.

\begin{figure}
 \begin{center}
\rotatebox{-90}{\includegraphics[width=6.5 cm]{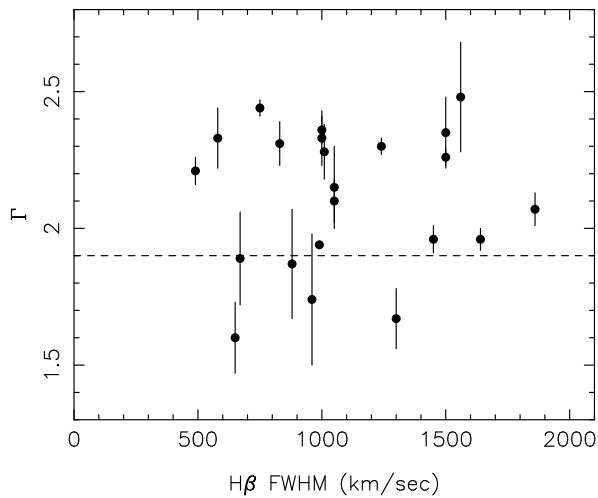}}
 \end{center}
 \caption{ The photon index of the hard continuum  plotted against 
the H$\beta$ line width for the NLS1 sources in the \asca\ sample.
The dotted line represents the mean
spectral index of the BLS1 sample of Nandra \et (1997). }
\end{figure}

In the present analysis we have fitted the soft excess in terms of a single 
temperature blackbody component, which should be a reasonable assumption, 
given the restricted \asca\ bandpass, if the 
emission resembles that expected from an accretion disc (e.g. Ross \et 1992). 
Indeed, the addition of a blackbody
component significantly improves the fit (at the $>99$\% confidence level
in the F-test) in 18 out of 22 objects. This is a much higher rate of incidence
of soft excesses
than appears to be the case for BLS1 samples (e.g. Turner \& Pounds 1989;
Reynolds 1997). The soft excess generally dominates over the hard power law 
at energies $^{<}_{\sim} 1.5$~keV, and presumably peaks in the extreme
ultraviolet. Even in the limited \asca\ bandpass the luminosity
of the soft excess is comparable to that of the hard power law and 
in three objects (1H~0707--495, IRAS~13224--3809 and PG~1404+226), the
blackbody component contains more luminosity than the power law in the
0.6--10~keV range, i.e., $L_{bb}/L_{pl} \geq 1$ (see Table~3). For these
three sources, extrapolation of the continuum spectrum down to
0.15~keV and up to 100 keV still leaves this ratio above unity.
The implication is that, at least in these NLS1s, the soft excess cannot 
solely be due to reprocessing of the hard X-ray continuum.  
(Note for these three objects the soft excess is still dominant even
after the inclusion of additional spectral features, see the Appendix.)

Two objects with strong soft excesses, namely
PHL~1092 and IRAS~13224--3809, appear to have particularly hard
underlying X-ray continua ($ \Gamma = 1.6-1.7 $, see Table~3).  This could
be used as an argument against models where the soft photons Compton scatter 
against the hot electrons in an accretion disk  corona to produce the
underlying X-ray hard spectrum.  However, an alternative view might be
that the relatively flat 2--10~keV spectra are the result of
strong reflection in these objects. Sensitive broad-band X-ray 
spectral measurements (extending well above the Compton peak)
would help distinguish between these two descriptions.

\subsection {Iron K$\alpha$ emission}

Emission lines at 6--7~keV are detected at a modestly significance
level in 9 objects with a marginal indication of such lines in 2 others. 
However, as noted earlier the line properties (energy, equivalent
width, intrinsic width) are in general only poorly constrained by the
\asca\ spectra. A point of interest, for follow-up when more sensitive
X-ray spectra are available, is the incidence of emission lines from
highly ionized gas (corresponding to Fe\textsc{xx} and above) in NLS1s,
for which the current data provide only a fleeting glimpse.

\subsection {Soft X-ray spectral features} 

The residuals to the best fitting continuum models show unusual
features below 2~keV in nine objects. In three sources
(NGC~4051, IRAS~17020+454 and IRAS~20181--224) we  are probably detecting
O\textsc{vii} and O\textsc{viii} edges  at 0.74 and 0.87 keV respectively, 
indicative of the presence of ``warm'' absorbing gas in our line of sight
to the nuclear continuum source. Of the other six objects, three 
(1H~0707--495, IRAS~13224--3809 and PG~1404+226) are most likely
affected by  absorption, whereas for the others (Ton~S180, PG~1244+026 and 
Ark~564) the interpretation is more ambiguous (see the Appendix). 
However, for the purpose of the present discussion we pursue the assumption 
adopted earlier, namely that all six sources exhibit anomalous absorption in 
the range 1.1--1.4~keV with the equivalent width of the absorption
feature being typically $\sim 100$~eV and its intrinsic width 
(when modelled as a pure Gaussian) ranging from 0.1--0.3 keV 
(see Table 4 and Fig. 4).

\begin{figure}
 \begin{center}
\rotatebox{-90}{\includegraphics[width=6.5 cm]{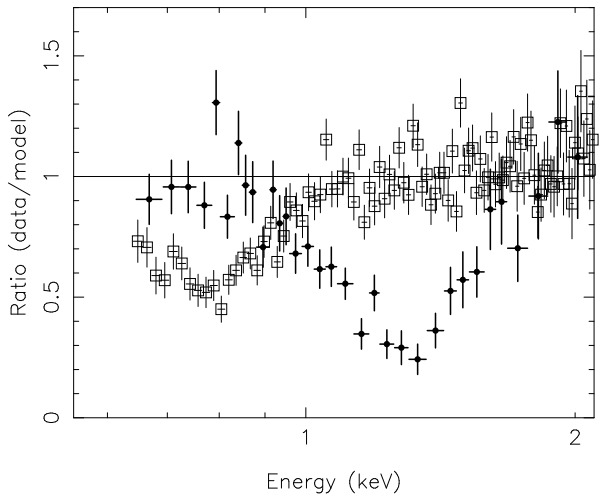}}
 \end{center}
 \caption{The ratio of the \asca\ data to the best-fit continuum model
for two sources,  IRAS~13224--3809 (filled circles) and IRAS~17020+454 
(open squares). The former source exhibits an anomalous absorption
feature near 1.2 keV and the latter probably a warm (\textsc{OVII}) edge near 0.74 keV.
Both spectra have been shifted into the source rest frame and
for clarity only SIS-0 data are shown.} 
\end{figure}

Leighly \et (1997b) have recently reported the detection of the same
anomalous absorption features in three sources and discussed 
possible physical origins of the effect. Specifically these authors
note that in order to identify the absorption with either an
O\textsc{vii} and/or 
an O\textsc{viii} edge, as predicted in the warm absorber scenario, then very
significant ({\it i.e.} relativistic) outflow velocities would be required 
in order to achieve the necessary blueshifting of the edge energies. 
An alternative model has been suggested 
by Nicastro, Fiore \& Matt (1999) who invoke resonant absorption lines in a
warm absorber to explain the absorption evident in the
spectrum of IRAS~13224-3809, in which case there is no requirement
for relativistic outflow. The basis of their model is that
a steep soft X-ray spectrum can produce a different ionization structure in
the warm absorber, with carbon and oxygen fully stripped. Absorption
is then produced by a complex
of absorption lines around 1--2~keV, made up of mainly iron L
resonance lines (see figure~4 of Nicastro \et 1999).
The total equivalent widths of these lines can be $\sim$100~eV, consistent 
with our results, although of course our assumption of a Gaussian
absorption profile is in this case an extremely crude approximation to 
the true underlying spectral form. 

\begin{figure}
 \begin{center}
\rotatebox{-90}{\includegraphics[width=6.5 cm]{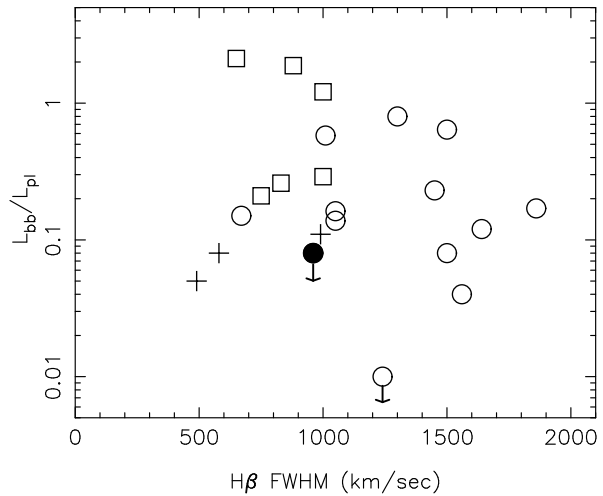}}
 \end{center}
 \caption{ The relative luminosity of the soft excess compared to the hard
power law (as given in Table~3) versus the H$\beta$ line width. Objects with no
evidence for absorption are marked by circles and those with normal warm
absorbers by crosses. The sources exhibiting anomalous 1.1--1.4~keV 
absorption are marked by squares and Mkn~507 (with possible cold absorption) is
marked with a filled circle. The arrows indicate upper limits on the strength of
the soft excess. }
\end{figure}

On the basis of the above arguments, it is at least plausible that
both the 0.7--0.9 keV and the 1.1--1.4 keV absorption arises due to
the presence of highly ionized gas along the line of sight to
the nuclear X-ray source. Interestingly the two different types of absorbers 
(described earlier as ``warm'' and ``anomalous'') occupy very different
regions of parameter space in Fig.~5, which shows the ratio of the luminosities
of the soft excess to the hard power law components ($L_{bb}/L_{pl}$)
versus the H$\beta$ line width.
The sources with
anomalous absorption clearly tend to have relatively high 
$L_{bb}/L_{pl}$ values whereas the three sources exhibiting ``normal'' warm 
absorption are found at the low end this range.
This is circumstantial evidence in support of the Nicastro et al. (1999) model,
in that they require a strong contribution from the soft excess 
in order to produce the necessary conditions for the dominance of
resonance over pure photoelectric absorption.
This point is given further weight if we also consider BLS1s in that
over 50\% of such sources show signs of absorption in the range 0.7--0.9~keV,
most likely due to oxygen edges, but relatively few possess  
conspicuous soft excesses, at least in the \asca\ bandpass 
(e.g., Reynolds 1997; George \et 1998).

A further interesting result illustrated by Fig.~6 is the fact
that all but one of the NLS1s with H$\beta$~FWHM~$\leq$~1000~km/s show 
evidence for some sort of absorption, either by ionized material or
(as in one case) by neutral gas. Conversely none of the NLS1s 
with H$\beta$~FWHM~$>$~1000~km/s exhibit signs of line-of-sight absorption.
One possibility is that this represents a very sharply
tuned geometry dependence. For example, if the absorbing material originates 
in an outflow from the disc, it would only be seen in low inclination systems
(for a discussion of the ``pole-on'' model for NLS1s, see Boller \et
1996 and references therein).  It is unclear, however, why such an incredibly
tight upper-limit should apply to the H$\beta$ line width in
such circumstances. Clearly further study of this effect using both
a larger sample of NLS1s and a more sophisticated modelling of
the putative absorption features is merited.

\section{Conclusions}

Our analysis of the \asca\ X-ray spectral data for a sample of 22
NLS1s has revealed a number of important characteristics of this
subclass of X-ray source. Absorption from {\it neutral} gas intrinsic to the 
host galaxy appears to be a very rare occurrence in NLS1. On the contrary most 
NLS1s exhibit a soft excess superposed on the hard power-law
continuum. Modelled as a blackbody component, this soft excess often contains 
a significant fraction of the X-ray luminosity in these sources, 
even in the restricted \asca\ bandpass. In the most extreme cases this
presents an argument against an origin for the soft excess in the 
reprocessing of the hard continuum. The underlying power-law spectra span a 
wide range of slopes  ($ \Gamma = 1.6-2.5 $) with a mean spectral index 
of $ 2.1 $, which is slightly steeper than the norm for BLS1s.
Iron K$\alpha$ emission lines are detected at a modestly significance
level in roughly half of the objects observed, but unfortunately
the line properties such as centroid energy, equivalent width
and intrinsic width are in general rather poorly constrained by the 
\asca\ data.

The X-ray spectra of NLS1s below 2~keV often exhibit additional features,
which in the current paper we have interpreted as largely due to
absorption by ionized material in the line of sight.
Three of the NLS1s show evidence for a absorption in the 0.7--0.9 keV range,
probably arising from O\textsc{vii} and O\textsc{viii} edges,  ``warm absorber'' features which 
are commonly observed in BLS1s. However, six NLS1s also show spectral features
which we take to be absorption in the 1.1--1.4 keV bandpass arising
from resonance absorption in highly ionized material.
A potentially very important result is that all objects showing
signs of absorption by highly ionized gas lie in the lower half
of the H$\beta$ line width distribution  ({\it i.e.}  the absorbed sources have
H$\beta$ FWHM $\leq 1000 \rm~km~s^{-1}$).  The combination of high 
effective area, good spectral resolution and extended bandwidth afforded by 
future missions such as XMM, AXAF and Astro--E, should allow unambiguous 
identification of the spectral features which appear to
characterise NLS1 galaxies and also confirm whether ionized absorption 
systems are preferentially observed in NLS1s with extremely narrow
H$\beta$ line widths.

\section*{Acknowledgments}

We thank Paul O'Brien and Ken Pounds for useful discussions, and the
referee, Neil Brandt, for his careful reading of the paper and comments.
This research made use of data obtained from the Leicester Database 
and Archive Service (LEDAS) at the Department of Physics and Astronomy, 
Leicester University, from the High Energy Astrophysics Science 
Archive Research Center (HEASARC), at the NASA Goddard Space
Flight Center and also from the NASA/IPAC Extragalactic Database (NED)
at JPL. SV acknowledges support from PPARC in the form of a research 
studentship.

\section {Appendix}

{\bf Some comments on the spectral fitting of individual objects}

\subsubsection*{Ton S180}

It is difficult to differentiate between absorption and emission at 
$\sim$1.2~keV in Ton~S180. A single absorption edge at a rest energy of 
E$=1.09\pm0.03$~keV with an optical depth $\tau=0.17\pm0.03$, yields
$\chi^{2}/{\nu}=1072/986$. A second edge in the spectrum (at
E$=1.42\pm0.06$ keV with $\tau=0.13\pm0.03$) improves the fit further
($\chi^{2}/{\nu}=1059/985$). There is some evidence for an additional edge 
at $\sim$0.73~keV ($\chi^{2}/{\nu}=1048/983$) presumably corresponding to
O\textsc{vii}. However SIS-1 seems to underestimate the flux compared to 
SIS-0 below 0.8~keV, so any features in the spectrum below
0.8 keV need to considered with caution.

A significant improvement over the soft excess fit is given
when a broad Gaussian emission component is added to the model near
1 keV. The best-fit parameters are line EW$=20\pm8$ eV, energy
E$=0.94\pm0.03$ keV and intrinsic width $\sigma=0.062\pm0.025$ keV,
for which $\chi^{2}/{\nu}=1055/985$. A \textsc{mekal} model also gives an 
acceptable fit to the data, yielding a plasma temperature of 
kT$=0.89\pm0.11$ keV ($\chi^{2}/{\nu}=1050/986$).

\subsubsection*{1H 0707-495}

An absorption feature definitely gives the best fit in 1H~0707--495.
A broad absorption line improves the fit significantly
($\chi^{2}/{\nu}=404/335$), as per Table~4.
An equally good $\chi^{2}$ can be obtained, but for 1 additional
free parameter, using two edges instead of a single Gaussian; the
first edge is at E=$1.09\pm0.03$~keV with $\tau=0.63\pm0.27$ and the second
at E=$0.90\pm0.03$~keV with $\tau=0.5\pm0.2$. 
Leighly \et (1997b) also find evidence for one or more
absorption features near 1~keV. The addition of a Gaussian emission line 
in the range 0.6--3~keV, instead of an absorption feature, does not provide 
any improvement in the fit.

\subsubsection*{NGC 4051}

Previous investigations of the X-ray spectrum of NGC~4051 have
suggested the existence of a warm absorber.
There is no evidence for additional spectral features in the first
observation (taken during the PV phase of the mission) but
the second (AO2) observation, which has much better signal/noise ratio,
does show significant features in the residuals between 0.7 and
0.9~keV. In this case we have modelled this apparent warm absorber
with 2 absorption edges, for the first edge E$=0.73\pm0.02$~keV and
$\tau=0.33\pm0.05$, for the second edge E=$0.93\pm0.03$~keV and
$\tau=0.19\pm0.04$. The resulting fit statistic is
$\chi^{2}/{\nu}=2010/1842$, an improvement of
$\Delta\chi^{2}=77$ over the model without absorption.
The results suggest a normal warm absorber
in NGC~4051, with the edges probably originating from O\textsc{vii} and
O\textsc{viii}. Our results are broadly consistent with those of Guainazzi \et
(1996), who consider the fitting of this observation in much greater
detail.

\subsubsection*{PG 1244+026}

Emission and absorption components give comparable fits in 
PG~1244+026. The addition of a broad Gaussian absorption line significantly
improves the fit  ($\chi^{2}/{\nu}=601/532$), as per Table~4. 
A single edge at an energy of E=$1.18\pm0.03$~keV, gives
a better fit ($\chi^{2}/{\nu}=591/533$). Adding another edge further
improves the  fit; with two edges (at E=$1.16\pm0.04$~keV and
E=$0.63^{+0.04}_{-0.45}$~keV) the fit obtained is $\chi^{2}/{\nu}=583/531$. 

A broad Gaussian emission line at an energy of E=$0.97\pm0.04$~keV also
significantly improves the fit ($\chi^{2}/{\nu}=589/532$), with an
equivalent width of EW=$36\pm11$~eV. 
A slightly better fit ($\chi^{2}/{\nu}=586/533$) is obtained with the
\textsc{mekal} model, with a plasma temperature of kT=$1.0\pm0.1$~keV.
Fiore \et (1998) also find evidence for either an
absorption feature at  $\sim$1.2~keV or an emission feature at 0.9~keV.

\subsubsection*{IRAS 13224-3809}

Absorption components give a slightly better fit than emission components.
The addition of a single absorption edge to the model improves the fit
significantly ($\chi^{2}/{\nu}=478/424$). The best fit edge energy is
E=$1.11\pm0.04$~keV. A second edge improves the fit further 
($\chi^{2}/{\nu}=467/422$), with edge energies at E=$1.00\pm0.04$ and
$1.18\pm0.03$~keV and optical depths of $\tau=0.73$ and 1.15
respectively. This fit also alters the continuum parameters slightly:
$\Gamma=1.84\pm0.07$ and kT=$163\pm9$~eV.
A broad Gaussian absorption line gives a slightly better fit
($\chi^{2}/{\nu}=459/423$), with parameters as in Table~4.

A broad Gaussian emission line at $0.74\pm0.05$~keV gives a slightly worse 
fit ($\chi^{2}/{\nu}=470/423$), as does the \textsc{mekal} model 
($\chi^{2}/{\nu}=480/424$) with a plasma temperature of
kT=$0.63\pm0.05$~keV. These results are broadly consistent with those of 
Leighly \et (1997b).

\subsubsection*{PG 1404+226}

Significant evidence is found for an absorption feature in PG 1404+226. The
addition of a single edge to the best-fitting continuum model (PL+BB)
yields $\chi^{2}/{\nu}=172/165$, an improvement of
$\Delta\chi^{2}=29$, with an edge energy of E$=1.07\pm0.03$~keV and
optical depth $\tau=0.84\pm0.28$, consistent with the results of Leighly
{\it et al.} (1997b). A slight improvement 
($\Delta\chi^{2}=8.1$) is also achieved by adding a second edge, although the
edge energy of E$=2.4\pm0.3$~keV seems rather high. The 1 keV feature can also
be modelled with a broad Gaussian absorption line, giving
$\chi^{2}/{\nu}=170/164$ with fit parameters as in Table~4. 

An emission model (Gaussian line) is unacceptable as the broad
emission feature essentially models the continuum soft
excess and gives an unacceptably large equivalent width.

\subsubsection*{IRAS 17020+454}

The residuals from a simple power law plus Galactic absorption fit show
clear signs of absorption in IRAS~17020+454 (see Figure~2). 
The addition of an edge at an energy of E=$0.71\pm0.02$ keVwith a
depth $\tau=0.91\pm0.22$, improves the fit
significantly($\chi^{2}/{\nu}=979/971$). A 
second edge at E=$1.15\pm0.04$~keV with a depth $\tau=0.20\pm0.08$
further improves the fit ($\chi^{2}/{\nu}=965/969$). 
A broad Gaussian absorption line gives a comparable fit as detailed 
in Table~4.

A narrow Gaussian emission line at an energy of E=$1.06\pm0.02$ 
with an equivalent width of EW=$23^{+5}_{-7}$~eV gives a
worse fit than an absorption line ($\chi^{2}/{\nu}=1006/971$).
However, the inclusion of an edge at $0.72\pm0.02$~keV, as well as the 
emission line, does lead to a net improvement in the fit 
($\chi^{2}/{\nu}=962/969$). 

In agreement with Leighly \et (1997a) we identify the 0.7~keV feature
with an absorption edge from O\textsc{vii}. Komossa \& Bade (1998) 
present a \rosat\ spectrum of IRAS~17020+454 and also find evidence for a 
warm absorber, albeit a dusty one.

\subsubsection*{IRAS 20181-224}

There may be an emission or absorption feature in the spectrum of
IRAS~20181-224, but the improvement in the fit upon adding either is rather 
small. A broad absorption line gives a slight improvement in the fit, as
noted in Table~4, but adding one or two edges in the range 0.6--3~keV gives 
a significantly a worse fit. 

A narrow Gaussian emission line at $1.11\pm0.05$~keV also slightly improves 
the fit ($\chi^{2}/{\nu}=438/387$). The \textsc{mekal} model
gives a similar result with a plasma temperature of kT=$1.7^{+0.6}_{-0.3}$~keV.

\subsubsection*{Ark 564}

Comparable fits are obtained with either a broad emission feature at
$\sim$1~keV or two absorption components. A single absorption edge at
E=$1.23\pm0.02$ keV gives $\chi^{2}/{\nu}=1862/1565$ whereas two edges
improves the fit to $\chi^{2}/{\nu}=1819/1563$ (the edge energies being
E=$1.14\pm0.03$~keV and $1.37\pm0.04$~keV with $\tau=0.18\pm0.04$ and
$0.15\pm0.03$, respectively). This is consistent with the Brandt \et
(1994)  modelling of the \rosat\ spectrum of Ark~564 which employed a deep
absorption edge at 1.15~keV. The broad Gaussian absorption line model
detailed in Table~4 is clearly not the best representation of the data
($\chi^{2}/{\nu}=1854/1564$). 

A broad emission line provides a slight improvement over two edges 
($\chi^{2}/{\nu}=1815/1564$) with E=$0.99\pm0.01$~keV, $\sigma = 0.10
 \pm 0.03$~keV and EW=$39^{+8}_{-4}$~eV.  The \textsc{mekal} model gives a 
worse fit ($\chi^{2}/{\nu}=1830/1565$), with a plasma temperature of
kT=$1.10^{+0.10}_{-0.05}$~keV. 
 
The SIS and GIS detectors systematically diverge below 0.9~keV.
There are also signs that the two SIS detectors are diverging at
$\sim$0.6~keV with SIS-1 seeming to underestimate the flux
compared with SIS-0. These problems make it difficult to establish the
nature of the soft X-ray spectral features in Ark~564.  

\bsp

\label{lastpage}


\begin{thebibliography}{99}

\bibitem{1} Brandt, W. N., Fabian, A. C., Nandra, K., Reynolds, C. S.,
Brinkmann, W.\ 1994, MNRAS, 271, 958
\bibitem{2} Brandt, W. N.\ 1996, Ph.D. Thesis
\bibitem{3} Brandt, W. N., Mathur, S., Elvis, M.\ 1997, MNRAS, 285, L25
\bibitem{4} Burke, B. E., Mountain, R. W., Harrison, D. C., Bautz,
M. W., Doty, J. P., Ricker, G. R., Daniels, P. J.\ 1991, IEEE Trans,
ED-38, 1069
\bibitem{5} Boller, Th., Brandt, W. N., Fink H. H.\ 1996, A\&A 305, 53
\bibitem{6} Comastri, A., \et 1998, A\&A, 333, 31
\bibitem{7} Czerny, B., Elvis, M. \ 1987, ApJ, 321, 305
\bibitem{8} Dickey J. M., Lockman F. J.\ 1990, ARA\&A 28, 215
\bibitem{9} Fabian, A. C.\ \et 1989, MNRAS, 238, 729
\bibitem{10} Fiore. F., \et 1998, MNRAS, 298, 103
\bibitem{11} George, I. M., Fabian. A.\ 1991, MNRAS, 249, 352
\bibitem{12} George, I. M., Turner, T. J., Netzer, H., Nandra, K.,
Mushotzky, R. F., Yaqoob, T.\ 1998, ApJS, 114, 73
\bibitem{13} Grupe, D.\ 1996, Ph.D. Thesis
\bibitem{14} Guainazzi, M., Mihara, T., Otani, C., Matsuoka, M.\ 1996,
PASJ, 48, 781
\bibitem{15} Iwasawa, K., Brandt, W.N., Fabian, A.C. \  1998, MNRAS, 293, 251
\bibitem{16} Leighly, K. M., Kay, L. E., Wills, B. J., Wills, D.,
Grupe, D.\ 1997a, ApJ, 489, L137
\bibitem{17} Leighly, K., Mushotzky, R., Nandra, K., Forster., K.\
1997b, ApJ, 489, L25
\bibitem{18} Kaastra, J. S., Mewe, R.\ 1993, A\&AS, 97, 443
\bibitem{19} Kohmura, Y., \et 1993, Proc SPIE, 2006, 78
\bibitem{20} Komossa, S., Bade, N.\ 1998, A\&A, 331, L49
\bibitem{21} Maraschi, L., Haardt, F. \ 1997,  In IAU Colloquium 163,
ASP Conference Series, ed. D. T. Wickramasinghe, G. V. Bicknell and 
L. Ferrario, 121,101
\bibitem{22} Matt, G., Fabian, A., Ross, R.\ 1993, MNRAS, 262, 179
\bibitem{23} Matt, G., Perola, G.C., Piro, L. \ 1991, A\&A, 247, 25
\bibitem{24} Mushotzky, R.F., Done, C., Pounds, K.A. \ 1993, ARA\&A, 31, 717
\bibitem{25} Nandra, K., Pounds, K. A.\ 1994, MNRAS, 268, 405
\bibitem{26} Nandra, K., George., I. M., Mushotzky, R. F., Turner,
T. J., Yaqoob, T.\ 1997, ApJ, 477, 602 
\bibitem{27} Nicastro, F., Fiore, F., Matt, G.\ 1999, ApJ, in press
\bibitem{28} Osterbrock, D. E., Pogge, R.\ 1985, ApJ 297, 166
\bibitem{29} Pounds, K., Done, C., Osborne, J.\ 1995, MNRAS, 277, L5
\bibitem{30} Reynolds, C. S.\ 1997, MNRAS, 286, 513
\bibitem{31} Ross, R., Fabian, A., Mineshige, S.\ 1992, MNRAS, 258, 189
\bibitem{32} Turner., T., Pounds, K.\ 1989, MNRAS, 240, 833
\bibitem{33} Wang, T., Brinkmann, W., Bergeron, J.\ 1996, A\&A, 309, 81
\bibitem{34} Zheng, W., O'Brien, P. T.\ 1990, ApJ, 353, 433

\end{thebibliography}
\end{document}